\documentclass[epsfig,psfig,aps,twocolumn,prb,showpacs]{revtex4-1}
\usepackage[colorlinks=true,citecolor=red,urlcolor=blue]{hyperref}
\usepackage{amsfonts,amssymb}
\usepackage{amsmath}
\usepackage{natbib} 
\usepackage{graphicx}
\usepackage{epsfig}
\DeclareMathOperator{\sgn}{sgn}

\begin{document}
\draft
\title
{
Electric field-induced valley degeneracy lifting in uniaxial strained graphene: evidence from magnetophonon resonance\\
}
\author
{
Mohamed Assili$^1$, Sonia Haddad$^1$ and Woun Kang$^2$
}
\address{
$^1$Laboratoire de Physique de la Mati\`ere Condens\'ee, D\'epartement de Physique,
Facult\'e des Sciences de Tunis, Universit\'e Tunis El Manar, Campus Universitaire 1060 Tunis, Tunisia\\
$^2$Department of Physics, Ewha Womans University, Seoul 120-750, Korea
}
%
%
\begin{abstract}
A double peak structure in the magneto-phonon resonance (MPR) spectrum of uniaxial strained graphene, under crossed electric and magnetic fields, is predicted. We focus on the $\Gamma$ point optical phonon modes coupled to the inter-Landau level transitions $0 \leftrightarrows \pm 1$ where MPR is expected to be more pronounced at high magnetic field. We derive the frequency shifts and the broadenings of the longitudinal (LO) and transverse (TO) optical phonon modes taking into account the effect of the strain modified electronic spectrum on the electron-phonon coupling. We show that the MPR line for a given phonon mode acquires a double peak structure originating from the two-fold valley degeneracy lifting. The latter is due to the different Landau level spacings in the two Dirac valleys resulting from the simultaneous action of the inplane electric field and the strain induced Dirac cone tilt.
We discuss the role of some key parameters such as disorder, strain, doping and electric field amplitude on the emergence of the double peak structure.
\end{abstract}
\maketitle

\section{Introduction}
The discovery of the anomalous quantum Hall effect (QHE) in graphene \cite{Geim} was the fingerprint of the presence of two-dimensional (2D) relativistic-like electrons, so called
Dirac electrons, which are at the origin of the striking properties of graphene.\

In the presence of a magnetic field perpendicular to the graphene sheet, the electron spectrum transforms into discrete
Landau levels (LL) with typical energy $\varepsilon_n=\sgn(n)\frac{\hbar v_F}{l_B}\sqrt{2|n|}$ where $n$ is an integer, $v_F$ is the Fermi velocity,
$l_B=\sqrt{\frac{\hbar}{eB}}$ is the magnetic length and $\sgn(n)=+(-)$ in the conduction (valence) band \cite{CastroRev,MarkRev}.\

To study the possible inter-Landau level excitations, magneto-Raman spectroscopy has proven to be a good probe which is remarkably sensitive
to the electron-phonon coupling \cite{Faugeras09,Goler}.
Such coupling in graphene has been, recently, a hot topic of study \cite{CastroRev}. The case of the center-zone doubly degenerate $E_{2g}$ optical phonon modes
and the corresponding Raman G band are found to be of particular interest \cite{CastroRev,Dresselhaus,Yan07,castro2007}.
By tuning the magnetic field, the G band shows a rich structure characterized by anticrossings known as magneto-phonon resonance (MPR) effect
\cite{Ando,MarkPRL}.
The latter occurs when the energy between two LL matches that of the G band phonon.\
The anticrossing behavior of MPR was first predicted by Ando \cite{Ando} who found that the frequency shifts
and the broadenings of the $\Gamma$ point optical phonon modes, in graphene, are marked by a singular behavior which can be tuned by doping.
Goerbig {\it et al.}\cite{MarkPRL} proposed that the fine structure resulting from the anticrossings of the MPR is expected
to split due to distinguishable behavior of circularly polarized phonons.
The properties of the split fine structure are found to be filling factor dependent\cite{MarkPRL}.
This dependence was proposed as a probe to measure the strength of the electron-phonon coupling and to resolve the phonon polarization \cite{MarkPRL}.\

Several magneto-Raman measurements have given evidence of the predicted fine structure properties of the MPR of the graphene G band
\cite{Faugeras09,Yan,Faugeras11,Kim12,Kuhne12,Kossacki,Qiu,Kim13} and in
particular its filling factor and light polarization dependences. The most pronounced effect of MPR was found, at high magnetic field, for the inter-LL transitions $0\rightarrow 1$ and $-1\rightarrow 0$ \cite{MarkPRL,Kossacki,Kim13}.
A fine structure of the G band was also observed in non-resonant regime\cite{Remi}.\

Recently, the filling factor dependence of the MPR was studied by controlling the position of the Fermi level in gated graphene \cite{Nano14,Shen}.\

Some puzzling features of the measured MPR structures remain unsolved \cite{Kossacki,Kim13}. In particular, the appearance of an additional component in the MPR structure \cite{Kim13} is still not clearly understood. Spatially inhomogeneous carrier densities and strain were proposed to be at the origin of this component \cite{Kim13}.

The behavior of the MPR in graphene under uniaxial strain was theoretically studied in Ref.~\onlinecite{Kashuba13}
taking into account the effect of the strain on the optical phonon spectrum.
The interplay between strain and MPR gives rise to a Raman line splitting and a light polarization preferences\cite{Kashuba13}.
In Ref.~\onlinecite{Kashuba13}, the authors did not consider the strain-induced modification of  the electron-phonon coupling.
They argued that, considering a linear electronic spectrum with strain induced anisotropic Dirac velocities, may simply result
in a weak renormalization of the magnetic field \cite{Kashuba13}.\

However, it has been shown that the electron-phonon interaction is sensitive to the modified electronic spectrum due to the strain \cite{Assili}.
In particular, the observed light polarization dependence of the G band in strained graphene was ascribed to the strain induced anisotropy
of the electronic dispersion \cite{Mohiuddin,Assili}.

The natural question which arises is how the MPR of the center zone optical phonon modes is modified if one takes into account the effect of the strain on the
electron-phonon interaction.

Under strain, the Dirac valleys of the honeycomb lattice are no longer located at the $K$ and $K'$ points of the Brillouin zone \cite{MarkRev}.
They are shifted away, depending on the strain amplitude. Moreover, the Dirac cones, which are isotropic in undeformed graphene, become tilted.\

A striking valley dependence of LL energy in graphene 
under crossed electric and magnetic fields was predicted by Lukose {\it et al.} \cite{Lukose}. 
The authors showed that, for a uniform electric field applied along the graphene sheet, the spacing between LL decreases and may eventually
vanish leading to a collapse of LL by increasing the electric field amplitude. 
This behavior is substantially different from that expected for a non-relativistic two dimensional electron
gas where the LL spacing is found to be independent of the electric filed \cite{Lukose}. \

Goerbig {\it et al.} \cite{Mark09} found that crossed electric and magnetic fields can induce a lifting of the twofold valley 
degeneracy of LL in tilted Dirac cones systems as in the organic conductor $\alpha$-(BEDT-TTF)$_2$I$_3$ where BEDT-TTF stands 
for the organic molecule bis(ethylenedithio)-tetrathiafulvalene. This salt is a layered material exhibiting Dirac-like 
electrons under pressure or uniaxial strain\cite{Kajita}.

Goerbig {\it et al.} \cite{Mark09} showed that, in the presence of a transverse magnetic field, 
the inplane electric field gives rise to an additional effective tilt term  which is valley dependent. 
The authors proposed to probe such valley filtering effect by infrared transmission spectroscopy.\\

Valley degeneracy lifting was recently investigated in graphene. Several ways were proposed to control the valley polarization \cite{valley}
such as introduction of topological line defects \cite{linedef} and strain engineering \cite{valleystrain}. The latter is expected to generate
pseudomagnetic fields with opposite sign in the two valleys. It is worth noting that the theory of the 
strain induced pseudomagnetic field and the underlying approximations were recently discussed in Ref.~\onlinecite{Ramezani}.\newline

A strained graphene barrier with artificially induced electron mass and spin-orbit coupling (SOC) was proposed  
as a possible spin-valley filter \cite{Grujic}. The electron mass and the SOC are expected to result from adatoms or substrate 
deposition \cite{Grujic}.

The valleytronics in Dirac electron-like systems has emerged as a possible way to encode carrier information as in spintronics.
One needs to remove the valley degeneracy to filter the electron of one valley \cite{valley}.\

In bismuth, in which the Dirac Hamiltonian was first introduced to describe the ``Dirac electrons'' in a condensed matter \cite{Wolff64}, valley degeneracy among three equivalent electron pockets was spontaneously removed at low temperature and/or high magnetic field. Loss of threefold symmetry was reported both in the magneto-transport \cite{Zhu12} and in the thermodynamic properties \cite{Kuchler14}.\\

In this paper, we consider a honeycomb lattice under uniaxial strain in the presence of crossed inplane electric field and a transverse magnetic field.
We investigate the behavior of the MPR spectrum taking into account the strain induced dependence of the electron-phonon interaction.
We focus on the center zone longitudinal optical (LO) and inplane transverse optical (TO) phonon modes corresponding to the G band in graphene.
We consider the coupling between the optical phonon modes and the electron-hole pairs resulting from inter-LL transitions
$0\leftrightarrows\pm 1$, where the effect of the MPR is more pronounced \cite{MarkPRL}.
The main result of this work is the possible occurrence of an electric field induced double peak structure in the MPR spectrum. This structure is the signature of the valley degeneracy lifting.
We also found that, considering strain modified electron spectrum, does not simply yield, as argued in Ref.~\onlinecite{Kashuba13}, to a weak renormalization
of the magnetic field but contributes to the fine structure of the phonon lines induced by the strain.\

The paper is organized as follows. In section II we derive the phonon self-energy in tilted Dirac cone systems under crossed electric and magnetic
fields. The shifts and the broadenings of the MPR modes are discussed in section III. A brief summary is given in section IV.

\section{Optical phonon self-energy}
It has been argued that the honeycomb lattice under uniaxial strain could be described by the minimal form of the generalized 2D Weyl
Hamiltonian\cite{MarkRev,katayama,Kobayashi07,morinari,Mark08} characterized by anisotropic
electronic velocities and a parameter measuring the tilt of the Dirac cones.
The tilt effect is not significant in deformed graphene contrary to the case of the organic conductor $\alpha$-(BEDT-TTF)$_2$I$_3$.\\

The 2D Weyl Hamiltonian can be written as \cite{MarkRev,Kobayashi07,morinari}:

\begin{eqnarray}
 H_{\xi}(\vec{k})=\xi\left( \vec{w}_0.\vec{k}\,\sigma^0+w_xk_x\sigma^x\right)+w_yk_y\sigma^y
\label{Helec}
\end{eqnarray}
where  $\xi=\pm$ is the valley index, $\vec{k}=(k_x,k_y)$ is the momentum vector. $\vec{w}_0=(w_{0x},w_{0y})$, $\sigma^0={1\!\!1}$, $\sigma^x$ and $\sigma^y$
are the $2\times 2$ Pauli matrices.\
$\vec{w}_0$ is responsible of the tilt of Dirac cones whereas $w_x$ and $w_y$ give rise to the anisotropy of the energy dispersion.\

Considering a uniaxial strain along the $y$ direction, the distance between first neighboring atoms along the $y$ axis changes from $a$ to $a^{\prime}=a+\delta a=a(1+\epsilon)$, where  $\epsilon=\frac{\delta a} a$ is the lattice deformation which measures the strain amplitude. $\epsilon$ is negative (positive) for compressive (tensile) deformation.\

The lattice basis changes from ($\vec{a}_1=\sqrt{3}a\vec{e}_x,\vec{a}_2=\frac{\sqrt{3}}2a\vec{e}_x+\frac32 a \vec{e}_y$)
to ($\vec{a}_1=\sqrt{3}a\vec{e}_x,\vec{a}_2=\frac{\sqrt{3}}2a\vec{e}_x+a\left(\frac32 +\epsilon\right)\vec{e}_y$) which corresponds to
a quinoid lattice \cite{Mark08}.\

In this case, the parameters of the generalized Weyl Hamiltonian (Eq.~\ref{Helec}) take the form
\begin{eqnarray}
w_x&=&\sqrt{3}a t \sin \theta, \, w_y=-ta\cos\theta+t^{\prime}a(1+\epsilon)\nonumber\\
w_{0x}&=&2\sqrt{3}a(t_{nnn}\sin 2\theta +t_{nnn}^{\prime}\sin \theta), \, \mathrm{and}\, w_{0y}=0
\label{wxy}
\end{eqnarray}
where $\theta=\arccos\left(-\frac{t^{\prime}}{2t}\right)$, $t$ and $t_{nnn}$ denote the hopping integrals, respectively,
to the first and to the second neighboring atoms along the bond directions which are not affected by the strain. $t^{\prime}$ ($t_{nnn}^{\prime}$)
is the hopping integral between first (second) neighbor atoms separated by the deformed lattice
distance $a^{\prime}=a(1+\epsilon)$ ($\|\vec{a}_2\|=\sqrt{3}a(1+\frac 23\epsilon)$). $t^{\prime}$ and $t_{nnn}^{\prime}$can be written as \cite{MarkRev}\
\begin{eqnarray}
t^{\prime}&=&t+\frac{\partial t}{\partial a}\delta a \nonumber\\
t^{\prime}_{nnn}&=&t_{nnn}+\frac{\partial t_{nnn}}{\partial a}\delta a
\end{eqnarray}
Given the Harrison's law\cite{MarkRev} $\frac{\partial t}{\partial a}=-\frac{2t}{a}$, $t^{\prime}$ becomes : $t^{\prime}=t(1-2\epsilon)$.

The electronic spectrum, corresponding to the Hamiltonian, given by Eq.~\ref{Helec} is :
\begin{eqnarray}
 \varepsilon_{\lambda}(\vec{k})=\vec{w}_0.\vec{k}+\lambda \sqrt{w_x^2k_x^2+w_y^2k_y^2}
\label{dispers}
\end{eqnarray}
where $\lambda=\pm 1$ is the band index.

For small strain amplitude, $w_x$, $w_y$ and $w_{0x}$ could be approximated by \cite{Mark08}:
\begin{eqnarray}
w_x&\simeq& \frac 32 at\left(1+\frac 23 \epsilon\right),
w_y\simeq \frac 32 at \left(1-\frac 43 \epsilon\right),\nonumber\\
w_{0x}&\simeq& 0.6 w_x\epsilon
\label{wxapp}
\end{eqnarray}

It is worth noting that we only consider the effect of a uniaxial strain and not a shear one,
to avoid complexity related to the contribution of Poisson ratio. The latter could be neglected as a first approximation \cite{Assili,Pereira}.
Moreover, we did not take into account the effect of the strain on the phonon dispersion to clearly identify the role of the electron-phonon interaction
in the strain induced MPR fine structure. These approximations were discussed in Ref.~\onlinecite{Assili}.\\

\subsection{Landau levels in tilted Dirac cones at zero electric field}

In the presence of a magnetic field transverse to the lattice plane $\vec{B}=B\vec{e}_z$ and in the Landau gauge $\vec{A}=(0,Bx,0)$,
the electronic spectrum splits into Landau levels (LL) given by:\cite{Himura}
\begin{eqnarray}
 \varepsilon_n=\sgn(n)\frac{\sqrt{w_xw_y}}{l_B}\sqrt{2\gamma^3|n|}
\end{eqnarray}
where $l_B=\sqrt{\frac{\hbar}{eB}}$ is the magnetic length, $\gamma=\sqrt{1-\tilde{w_0}^2}$ and $\tilde{w}_0$ is the
tilt parameter given by: 
\begin{eqnarray}
\tilde{w}_0=\sqrt{\left(\frac{w_{0x}}{w_x}\right)^2+\left(\frac{w_{0y}}{w_y}\right)^2}.
\label{tiltw0}
\end{eqnarray}

The corresponding eigenfunctions are of the form \cite{Ando,Himura}
\begin{eqnarray}
\langle \vec{r}| nX\rangle=F_{nX}(\vec{r})=\frac 1{\sqrt{L_y^{\prime}}}\exp({i \frac{Xy}{l_B^2}}) \phi_n^{\xi}(x)
\label{Fnx}
\end{eqnarray}
where $X=k_y l^2_B$ is the center of coordinates and $L_y^{\prime}$
is the crystal length along the strain direction $y$. $\phi_n^{\xi}(\vec{r})$
are given by\cite{Himura}:
\begin{widetext}
\begin{eqnarray}
\phi_n^{\xi}&=&\frac{1}{2\sqrt{1+\xi\gamma}}  \left(
\begin{array}{c}
-\xi \tilde{w}_0\mathrm{e}^{-i\xi \phi_t}\\
\xi+\gamma
\end{array}
\right)
f_{|n|}+
\frac{\sgn(n)}{2\sqrt{1+\xi\gamma}}  \left(
\begin{array}{c}
(1+\xi\gamma) \mathrm{e}^{-i\xi \phi_t}\\
-\tilde{w}_0
\end{array}
\right)
f_{|n|-1}
\label{phi_n}
\end{eqnarray}
\begin{eqnarray}
\phi_0^{\xi}&=&\frac{1}{\sqrt{2(1+\xi\gamma)}}  \left(
\begin{array}{c}
-\xi \tilde{w}_0\mathrm{e}^{-i\xi \phi_t}\\
\xi+\gamma
\end{array}
\right)
f_{0}
\label{phi_0}
\end{eqnarray}
\end{widetext}
where $\tan \phi_t=\frac{w_{0y}w_{x}}{w_{y}w_{0x}}$. In the present case, $\phi_t=0$ since $w_{0y}=0$ (Eq.~\ref{wxy}). \

The $f_{|n|}$ functions are given by \cite{Himura}
\begin{equation}
f_{|n|}(u)=\frac{(-1)^n}{\sqrt{\pi^{\frac 12}2^{|n|}|n|!l_B}}\exp{\left(-\frac{u^2}{2l_B^2}\right)}H_{|n|}\left(\frac{u}{l_B}\right)
\end{equation}
$H_{|n|}$ being the Hermite polynomials and 
$u=\sqrt{\gamma}\left(x-l^2_Bk_y\right)-\xi \sgn(n) \tilde{w}_0l_B\sqrt{2|n|}$ [Refs.~\onlinecite{Lukose},~\onlinecite{Himura}].\\

It is worth stressing, that we have derived Eqs.~\ref{phi_n} and \ref{phi_0} using the method proposed by Himura {\it et al.} \cite{Himura} 
However, the form of $\phi_n^{\xi}$ function (Eq.~\ref{phi_n}) is slightly different from that found
in Ref.~\onlinecite{Himura} where a $\sgn(n)$ was missing and a factor of $\xi+\gamma$ should be replaced by $1+\xi\gamma$.

\subsection{Landau levels in tilted Dirac cones under a uniform electric field}

The electron spectrum in tilted Dirac cone systems, in the presence of an inplane electric field $\vec{E}=E \vec{e}_y$ 
is characterized by an electric field dependent tilt vector \cite{Mark09}
\begin{eqnarray}
\vec{w}_{\xi}(E)\equiv \left(w_{\xi x},w_{\xi y} \right)=\vec{w}_0-\xi \hbar \frac{\vec{E}\times \vec{B}}{B^2}
\end{eqnarray}
with $\vec{w}_0=\left( w_{0 x},w_{0 y}=0\right)$ is the tilt vector in the absence of the electric field (Eq.~\ref{wxy}).\

The tilt parameter is given by \cite{Mark09}:
\begin{eqnarray}
\tilde{w}_{\xi}(E)=\sqrt{\left(\frac{w_{\xi x}}{w_x}\right)^2+\left(\frac{w_{\xi y}}{w_y}\right)^2}
\label{tiltE}
\end{eqnarray}
The tilt angle under electric field is written as \cite{Mark09} $\tan \Phi^{\xi}_t(E)=\frac{w_{\xi y}w_x}{w_{\xi x}w_y}$.\

Under a uniform electric field $\vec{E}=E\vec{e}_y$, the tilt parameter, given by Eq. \ref{tiltw0}, becomes:
\begin{eqnarray}
\tilde{w}_{\xi}(E)=\frac 1{w_x}|w_{0 x}-\xi \hbar \frac{E}{B}|
\label{tiltEw}
\end{eqnarray}
The LL spectrum is given by \cite{Lukose,Mark09}
\begin{eqnarray}
\epsilon^{\xi}_n(E)=\sgn(n) \frac{\sqrt{w_xw_y}}{l_B}\left[1-\tilde{w}^2_{\xi}(E)\right]^{\frac 34} \sqrt{2|n|}+\frac{\hbar E}{B} k_x\nonumber\\
\label{LL-E}
\end{eqnarray}
The spacing between LL is then valley dependent: it is larger in $\xi=+$ ($\xi=-$) valley 
for a tensile (compressive) deformation as we shall show in the next section \cite{Mark09}.\

The LL eigenfunctions $\phi^{\xi}_n(E)$ have the same form as those given by Eqs. \ref{phi_n} and \ref{phi_0} but with valley dependent tilt parameters:\cite{Mark09} $\gamma^{\xi}(E)=\sqrt{1-\tilde{w}^2_{\xi}(E)}$ where $\tilde{w}_{\xi}(E)$ is given by Eq.~\ref{tiltE}.

\subsection{Electron-phonon interaction Hamiltonian}
The electron-phonon Hamiltonian in graphene can be written as \cite{ando2006}:
\begin{eqnarray}
H_{int}^{\xi}=-\sqrt{\frac{\hbar}{NM}}\frac{\beta}{a^2}\sum_{\vec{q},\mu}\frac
1{\sqrt{\omega_{0,\mu}}}V_{\mu}^{\xi}(\vec{q})\mathrm{e}^{i\vec{q}.\vec{r}}\left(b_{\vec{q},\mu}+b^{\dagger}_{-\vec{q},\mu}\right)\nonumber\\
\label{Hint}
\end{eqnarray}
where $M$ is the mass of the carbon atom, $b^{\dagger}_{\vec{q},\mu}$ ($b_{\vec{q},\mu}$) is the creation (annihilation) operator of phonon with wave vector $\vec{q}=(q_x,q_y)$.
$\beta=-\frac{d\ln t}{d\ln a}=-\frac a t \frac{\partial t}{\partial a}$, $\omega_{0\mu}$ is the $\Gamma$ point optical phonon frequency of
the mode $\mu$ in the absence of electron-phonon interaction. $\mu=L$ ($\mu=T$) for LO (TO) phonon mode.\

In undeformed graphene $\omega_{0T}=\omega_{0L}=\omega_{0}$.
This degeneracy is lifted in the strained graphene due to the symmetry breaking. However, we assume, for simplicity, that $\omega_{0T}\simeq\omega_{0L}\simeq\omega_{0}$.
This turns out to neglect the effect of the strain on the phonon dispersion since we are interested in the role of the electronic dispersion on the magnetophonons \cite{Assili}.\\

In Eq.~\ref{Hint}, the matrices $V_{\mu}^{\xi}(\vec{q})$ near the Dirac points $D$ ($\xi=+$) and $D^{\prime}$ ($\xi=-$) are given by\cite{Assili}.:
\begin{widetext}
\begin{eqnarray}
V_L^{\xi}(\vec{q})&=&\sqrt{w_xw_y^{\prime}}\left(
\begin{array}{cc}
0& i\frac{\sin \varphi(\vec{q})}{\alpha^{\prime}}-\xi\alpha^{\prime}\cos \varphi(\vec{q})\\
i\frac{\sin \varphi(\vec{q})}{\alpha^{\prime}}+\xi\alpha^{\prime}\cos \varphi(\vec{q})&0\\
\end{array}
\right)\nonumber\\
V_T{\xi}(\vec{q})&=&\sqrt{w_xw_y^{\prime}}\left(
\begin{array}{cc}
0& i\frac{\cos \varphi(\vec{q})}{\alpha^{\prime}}+\xi\alpha^{\prime}\sin \varphi(\vec{q})\\
i\frac{\cos \varphi(\vec{q})}{\alpha^{\prime}}-\xi\alpha^{\prime}\sin \varphi(\vec{q})&0\\
\end{array}
\right)\nonumber\\
\label{V}
\end{eqnarray}
 \end{widetext}
where $\varphi(\vec{q})$ is the phonon angle $\tan\varphi(\vec{q})=\frac{q_y}{q_x}$,
$w_y^{\prime}=w_y-2\epsilon t^{\prime}a(1+\epsilon)\simeq w_y(1-\frac 43 \epsilon)$ and $
\alpha^{\prime}=\sqrt{w_x/w_y^{\prime}}$.

\subsection{Optical phonon self-energy}
The total self-energy $\Pi_{\mu}$ of a given optical phonon mode ($\mu=$LO, TO) is due to the interaction of phonons with electrons in both valleys ($\xi=\pm$), it is written as \cite{Ando}
\begin{eqnarray}
 \Pi_{\mu}=\sum_{\xi=\pm}\Pi_{\mu}^{\xi}
\end{eqnarray}

The $\Gamma$ point optical phonon self-energy of a mode $\mu$, due to interactions with electron-hole pairs near the valley $\xi$, is given by \cite{Ando}
\begin{widetext}
\begin{eqnarray}
\Pi_{\mu} ^{\xi}(\vec{q}\rightarrow \vec{0},\omega)=-g_s
\frac{\hbar S^{\prime}}{NM\omega_0}
\left(\frac{\beta}{a^2}\right)^2\sum_{n,n^{\prime}}\sum_X
|\langle nX|V_{\mu}^{\xi}(\vec{q})|n^{\prime}X\rangle|^2
\frac{f\left(\varepsilon^{\xi}_{n}(E)\right)-
f\left(\varepsilon^{\xi}_{n^{\prime}}(E)\right)}
{\hbar \omega+\varepsilon^{\xi}_{n^{\prime}}(E)
-\varepsilon^{\xi}_{n}(E)+i\delta}
\label{self}
\end{eqnarray}
\end{widetext}
where we considered the valley dependent LL spectrum under electric field $\vec{E}=E\vec{e}_y$ (Eq.~\ref{LL-E}).
The functions $\langle \vec{r}| nX\rangle$ are the LL eigenfunctions given by Eq.~\ref{Fnx}, $f(x)$ is the Fermi-Dirac distribution, 
$g_s$ is the spin degeneracy and $\delta=\frac{\hbar}{\tau}$, $\tau$ being the scattering time.\

The sum over $X$ gives rise to the degeneracy of the LL $g_L=\frac{S^{\prime}}{2\pi l_B^2}$, where $S^{\prime}$ is
the surface of the strained graphene which can be expressed in term of the undeformed
surface $S$ as $S^{\prime}=N \|\vec{a}_1\times \vec{a}_2\| \simeq S\left(1+\frac 23 \epsilon\right) $. \\

The self-energy Eq.~\ref{self} exhibits a resonance behavior if the following condition is fulfilled
\begin{eqnarray}
\hbar \omega_0 +\varepsilon^{\xi}_{n^{\prime}}(E)
-\varepsilon^{\xi}_{n}(E)\simeq 0
\label{condition}
\end{eqnarray}
where $\hbar\omega_0=0.196$ eV is the bare optical phonon energy.\

Since the $\Gamma$ phonon modes induce a vertical inter-LL transitions, the momentum dependent term $\frac{\hbar E}B k_x$ 
in the energy spectrum Eq.~\ref{LL-E}, responsible of the inclination of the LL, can be omitted.
The inclination effect is not relevant since, in a given valley, the LL spacing is constant.
Moreover, the contribution of such effect compared to the resonance energy can be estimated by calculating 
the ratio $\frac{\Delta \varepsilon}{\hbar\omega_0}$ where $\Delta \varepsilon=\hbar k_0\frac E B$, $k_0$ 
being the momentum value corresponding to a resonant inter-LL $0 \leftrightarrows \pm 1$ transitions: 
$\hbar \omega_0 \sim \hbar v_F k_0$ where $v_F\sim 10^6$ m.s$^{-1}$ is the average Fermi velocity in graphene. We then obtain
\begin{eqnarray}
\frac{\Delta \varepsilon}{\hbar\omega_0} \sim \frac{E}{v_F B}
\end{eqnarray}

The MPR associated to inter-LL $0 \leftrightarrows \pm 1$ transitions, take place around $B$= 30 T \cite{Faugeras11,Kuhne12,Kossacki}. 
For an electric field of the order of $E\sim 10^6$ V/m, we have $\frac{\Delta \varepsilon}{\hbar\omega_0}\simeq 3\%$. 
The inclination of the LL can then be neglected in this case.

Using Eqs.~\ref{phi_n} and \ref{phi_0}, the matrix elements in Eq.~\ref{self} take the form:
\begin{widetext}
\begin{eqnarray}
|\langle nX|V_{L}^{\xi}(\vec{q})|n^{\prime}X\rangle|^2&=&w_xw_y^{\prime}
K_n^{\xi2}K_{n^{\prime}}^{\xi2}\left[ \delta_{|n^{\prime}|,|n|-1}+\delta_{|n|,|n^{\prime}|-1}\right]\times\nonumber\\
&&\left\lbrace \left( \frac{\sin\varphi}{\alpha^{\prime}}\right)^2
\left[ ( \xi+\gamma^{\xi}(E))^2+\tilde{w}_{\xi}^2(E)\right]^2+
\left({\alpha^{\prime}\cos\varphi}\right)^2
\left[ ( \xi+\gamma^{\xi}(E))^2-\tilde{w}_{\xi}^2(E)\right]^2\right\rbrace \nonumber\\
|\langle nX|V_{T}^{\xi}(\vec{q})|n^{\prime}X\rangle|^2&=&w_xw_y^{\prime}
K_n^{\xi2}K_{n^{\prime}}^{\xi2}\left[ \delta_{|n^{\prime}|,|n|-1}+\delta_{|n|,|n^{\prime}|-1}\right]\times \nonumber\\
&&\left\lbrace \left( \frac{\cos\varphi}{\alpha^{\prime}}\right)^2
\left[ ( \xi+\gamma^{\xi}(E))^2+\tilde{w}_{\xi}^2(E)\right]^2+
\left({\alpha^{\prime}\sin\varphi}\right)^2
\left[ ( \xi+\gamma^{\xi}(E))^2-\tilde{w}_{\xi}^2(E)\right]^2\right\rbrace
\label{matrice}
\end{eqnarray}
\end{widetext}
here $\varphi=\varphi(\vec{q})$ is the phonon angle, $K_{n\neq0}^{\xi}=\frac1{2\sqrt{1+\xi\gamma^{\xi}(E)}}$ and $K_{n=0}^{\xi}=\frac1{\sqrt{2(1+\xi\gamma^{\xi}(E))}}$.\

As mentioned in Refs.~\onlinecite{Ando},~\onlinecite{ando2006}, the contribution of $\omega=0$ should be subtracted from the phonon self-energy
to avoid a double counting problem. We then obtain:
\begin{widetext}
\begin{eqnarray}
\Pi_{\mu} ^{\xi}(\vec{q}\rightarrow \vec{0},\omega)=-C A_{\mu} ^{\xi}\sum_{\lambda,\lambda^{\prime}=\pm1}\sum_{n=0}C_{n} ^{\xi}\;
 \frac{2\left[ f\left(\lambda\varepsilon^{\xi}_{n}(E)\right)-
f\left(\lambda^{\prime}\varepsilon^{\xi}_{n+1}(E)\right)\right]\left( \lambda \varepsilon^{\xi}_{n}(E)-\lambda^{\prime}\varepsilon^{\xi}_{n+1}(E)\right) }
{\left( \hbar \omega+i\delta\right)^2- \left( \lambda \varepsilon^{\xi}_{n}(E)-\lambda^{\prime}
\varepsilon^{\xi}_{n+1}(E)\right)^2}
-\frac{1-\lambda\lambda^{\prime}}{\varepsilon^{\xi}_{n+1}(E)+\varepsilon^{\xi}_{n}(E)}\nonumber\\
\label{self2}
\end{eqnarray}
\end{widetext}

$C_{n\neq 0} ^{\xi}=\left( \frac 1{4(1+\xi\gamma^{\xi}(E))}\right) ^2$ and $C_{n=0} ^{\xi}= \frac 1{8(1+\xi\gamma^{\xi}(E))^2}$. 
The constant $C$ is given by:
\begin{eqnarray}
C&=&g_s36\sqrt{3}\left(\frac{\beta}{2}\right)^2\frac{\hbar}{2Ma^2\omega_0}\frac1{\hbar \omega_0}\nonumber\\
&\simeq& g_s36\sqrt{3}\left(\frac{\beta}{2}\right)^2\frac{\hbar}{2Ma^2\omega_0}\frac1{\hbar \omega_0},
\end{eqnarray}
here $a_0=a\sqrt{3}$ is the lattice parameter in the undeformed system. The coefficients $A_{\mu}^{\xi}$ are given by:
\begin{widetext}
\begin{eqnarray}
A_{L}^{\xi}&=&w_xw_y^{\prime}
\left\lbrace \left( \frac{\sin\varphi}{\alpha^{\prime}}\right)^2
\left[ ( \xi+\gamma^{\xi}(E))^2+\tilde{w}_{\xi}^2(E)\right]^2+
\left({\alpha^{\prime}\cos\varphi}\right)^2
\left[ ( \xi+\gamma^{\xi}(E))^2-\tilde{w}_{\xi}^2(E)\right]^2\right\rbrace \nonumber\\
A_{T}^{\xi}&=&w_xw_y^{\prime}
\left\lbrace \left( \frac{\cos\varphi}{\alpha^{\prime}}\right)^2
\left[ ( \xi+\gamma^{\xi}(E))^2+\tilde{w}_{\xi}^2(E)\right]^2+
\left({\alpha^{\prime}\sin\varphi}\right)^2
\left[ ( \xi+\gamma^{\xi}(E))^2-\tilde{w}_{\xi}^2(E)\right]^2\right\rbrace \nonumber\\
\label{matrice2}
\end{eqnarray}
\end{widetext}

Since these coefficients are valley dependent, the phonon frequency shifts $\Delta \omega_{\mu}$ and the corresponding broadenings $\Gamma_{\mu}$
are also expected to depend on the valleys and they are given by:
\begin{eqnarray}
\Delta\omega^{\xi}_{\mu}&=&\omega_{\mu}-\omega_0=\frac 1{\hbar} \Re\,\Pi^{\xi}_{\mu}(\vec{q},\omega_0)\nonumber\\
\Gamma^{\xi}_{\mu}&=&-\frac 1{\hbar} \Im\,\Pi^{\xi}_{\mu}(\vec{q},\omega_0)
\label{omega-Gamma}
\end{eqnarray}

Let us, first, focus on the zero electric field case. The resonance occurs, according to Eq.~\ref{self2}, 
when the phonon frequency is equal to the energy separation between two LL:
$\hbar \omega_0=\lambda \varepsilon_{n}-\lambda^{\prime}
\varepsilon_{n+1}$.\

For interband ($\lambda=-\lambda^{\prime}$) and intraband ($\lambda=\lambda^{\prime}$) transitions between LL, the resonance takes place, respectively, at magnetic energy $\hbar\omega_B^{n_{+}}$ and $\hbar\omega_B^{n_{-}}$
satisfying \cite{Ando}:
\begin{eqnarray}
\hbar\omega_B^{n_{\pm}}&=&\sqrt{w_xw_y}\frac{\sqrt{2\gamma^3}}{l_B}\left[ \sqrt{n+1}\pm\sqrt{n}\right]\nonumber\\
&=&\hbar \omega_B \frac {2}3 \frac{\sqrt{w_xw_y}}{at}\gamma^{\frac 32}\left[ \sqrt{n+1}\pm\sqrt{n}\right]
\label{energyB}
\end{eqnarray}
where $\hbar\omega_B=\frac 32 at \frac{\sqrt{2}}{l_B}$ [Ref.~\onlinecite{Ando}].\

In the case of uniaxial strained graphene, $\hbar\omega_B^{n_{\pm}}$ could be written, in the limit of small strain amplitude, as:
\begin{eqnarray}
\hbar\omega_B^{n_{\pm}}&\simeq& \hbar\omega_B\left(1-\frac 13\epsilon\right)\left( 1-\frac 34(0.6\epsilon)^2\right) )\left[ \sqrt{n+1}\pm\sqrt{n}\right]\nonumber\\
\label{energyBsim}
\end{eqnarray}
where we considered the approximate values of $w_x$, $w_y$ and $\tilde{w}_0$ at zero electric field (Eq.~\ref{wxapp}).\

Eq.~\ref{energyBsim} shows that the resonance takes place at higher (lower) magnetic field for tensile (compressive) deformation compared
to the undeformed case. According to Eq.~\ref {energyBsim}, and disregarding the strain modified phonon spectrum,
the electron-phonon interaction in strained graphene induces a shift in the anticrossing position of the magnetophonon spectrum. 
This shift is due to the strain modified electron-spectrum.
The shifted position of the MPR, compared to that in the undeformed lattice, is an
indication of the presence of deformation. The latter may accidentally occur during the fabrication 
procedure of the graphene sheet by exfoliation or chemical vapor deposition.\\

Let us now turn to the possible evidence of the valley filtering in the MPR spectrum.
According to Eqs.~\ref{self2} and \ref{matrice2}, the frequency shifts of a given phonon mode is valley dependent. However, at zero electric field, the resonance condition, which is responsible of the anticrossing structure of the MPR, does not depend on the Dirac valleys (Eq.~\ref{energyB}). One should, then, not expect in the absence of the electric field, any signature of the valley degeneracy lifting in the MPR spectrum. \

However, under an inplane electric field, the resonance condition is valley dependent (Eq.~\ref{condition}). 
The self-energy is then expected to have a singular behavior at two critical
magnetic energy $\hbar\omega_B^{\xi=+}$ and $\hbar\omega_B^{\xi=-}$ 
corresponding to the valleys $\xi=\pm$. If the LL spacing in the $\xi=+$ valley is larger than in the $\xi=-$ (Eq.~\ref{LL-E}), 
the resonance due to inter-LL transitions in $\xi=+$ valley will then appear before that associated to $\xi=-$, by increasing the magnetic field,
since $\hbar\omega_B^{\xi}$ is given by:
\begin{eqnarray}
\hbar\omega_0=\frac 23\hbar\omega_B^{\xi}  \frac{\sqrt{w_xw_y}}{at} \left[1-\tilde{w}^2_{\xi}(E)\right]^{\frac 3 4}
\left[\sqrt{n+1}+\sqrt{n}\right]\nonumber\\
\label{wBxi}
\end{eqnarray}
A double peak in the MPR lines may, then, appear in the presence of the inplane electric field.\\

It is worth stressing that, besides the phonon line splitting induced by valley degeneracy lifting,
one may expect a change in the MPR spectrum due to the electron tunneling between valleys. This effect takes place at high magnetic field or as
the merging of Dirac cones is approached \cite{Gilles}. 
However, such effect is not relevant in graphene since a large strain amplitude is required ($\sim 30\%$) to observe it.

Due to the electron intervalley tunneling the zero LL splits into two levels $0^+$ and $0^-$ separated by an energy gap \cite{Gilles,Assili13}
\begin{eqnarray}
\Delta E\propto\mathrm{exp}\left(-4\sqrt{2} \frac{\alpha^6}{\tilde{B}}\right)
\label{gap}
\end{eqnarray}
where $\alpha=\sqrt{\frac{w_x}{w_y}}\simeq 1+\epsilon$ and $\tilde{B}$ is a dimensionless parameter given by $\tilde{B}=\frac{Bea^2}{\hbar}=\frac B{B_0}$
with $B_0=\frac{\hbar}{ea^2}$.\

The intra-LL transitions $0^+ \longrightarrow 1$ and $-1 \longrightarrow 0^-$ will appear at different magnetic fields 
compared to the inter-LL transitions $0^- \longrightarrow 1$ and $-1 \longrightarrow 0^+$. In the absence of electron intervalley tunneling,
the $-1 \longrightarrow 0$ transition is forbidden for a totally filled $n=0$ level. The corresponding Raman line obtained 
with circularly polarized light disappears and only the line ascribed to $0 \longrightarrow 1$ transition persists at high magnetic field.\

In the presence of the intervalley tunneling, besides the disappearance of the $-1 \longrightarrow 0$ line, the intensity of the inter-LL $0^- \longrightarrow 1$ transition will be reduced since the intra-LL transition $0^+ \longrightarrow 1$ will no more contribute to the inter-LL line $0^- \longrightarrow 1$. The $0^+ \longrightarrow 1$ intra-LL line may emerge at higher field compared to the inter-LL.\\

In graphene, the distance between first neighboring atoms is $a=1.42$~\AA. For a critical compressive strain $\epsilon=30\%$, before sample cracking,
the anisotropy parameter is $\alpha\simeq 0.7$, which gives rise to $B_0\simeq 10^4$~T (Eq.~\ref{gap}).
A magnetic field of $10^3$ T is, then, needed to detect the change in the MPR spectrum of graphene due the valley electron tunneling. 
However, in the organic salt $\alpha$-(BEDT-TTF)$_2$I$_3$, the anisotropy parameter $\alpha$ could be of the order 
of $0.4$ and $B_0\simeq 660$~T ($a\simeq 10$~\AA). The changes in the MPR spectrum induced by the electron valley 
tunneling may then be observed in $\alpha$-(BEDT-TTF)$_2$I$_3$ at a field amplitude of $B\gtrsim 15$~T.

\section{Results and discussion}

We first focus on the case of uniaxial strained graphene in the absence of the electric field and then turn to the study of the electric field induced valley degeneracy lifting.

\subsection{MPR in uniaxial strained graphene at zero electric field}
We numerically derive the frequency shifts and the broadenings of the LO and the TO modes corresponding to $0\leftrightarrows\pm 1$ transition
(Eqs.~\ref{self2} and \ref{omega-Gamma}).\

We consider linearly polarized phonon modes along (LO) and perpendicular (TO) to the phonon momentum direction 
$\left(\cos \varphi(\vec{q}),\sin \varphi(\vec{q})\right)$ where $\varphi(\vec{q})$ is the phonon angle. We denote hereafter, $\varphi(\vec{q})=\phi$.
For $\phi=\frac{\pi} 2$, the LO mode is along the strain direction.\newline

\begin{figure}[hpbt]
\begin{center}
\vspace{0.5cm}

\includegraphics[width=0.9\columnwidth]{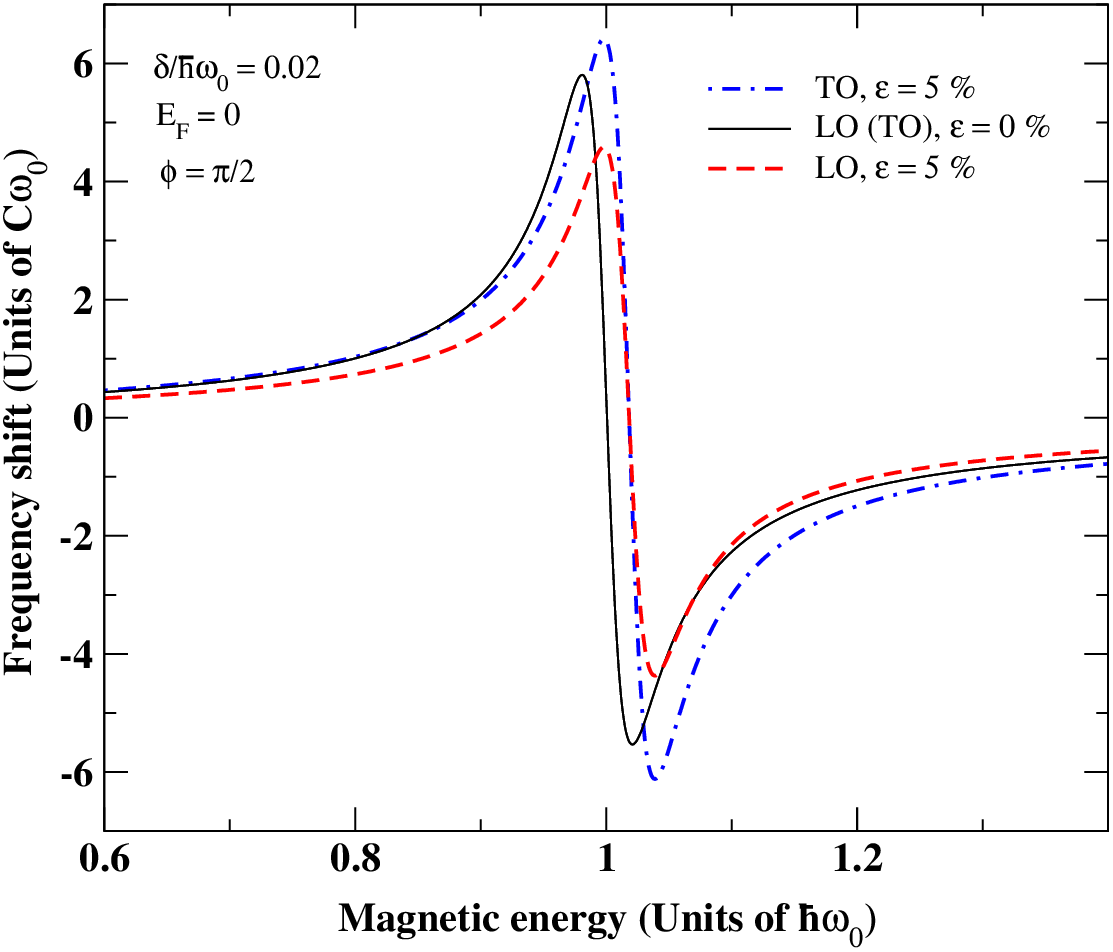}
\end{center}
\caption{Frequency shifts of LO and TO modes as a function of the magnetic energy $\hbar\omega_B$ in undoped graphene ($E_F=0$) for a disorder amount of $\frac 1{\tau\omega_0}=0.02$ and under a tensile strain $\epsilon=\frac{\delta a}a=5\%$. The LO mode is along the strain axis. The thin solid line is the result for the undeformed case.}
\label{shiftlow}
\end{figure}
Figure \ref{shiftlow} shows the frequency shifts of the LO and the TO modes in undeformed lattice and for a tensile deformation of $5\%$. The calculations are done in the undoped regime.\ 

As shown in Fig.\ref{shiftlow}, the strain splits the degenerate phonon line of the undeformed lattice into two lines corresponding to the two linearly
polarized LO and TO modes.
Kashuba and Fal'ko \cite{Kashuba13} showed that the Raman signal splits under strain but did not address the strain dependence 
of the anticrossing position.\

According to Fig.\ref{shiftlow}, the anticrossings marking the MPR, corresponding to the  $ 0\rightleftarrows \pm 1$, 
take place at slightly shifted magnetic field value compared to the undeformed case where MPR occurs at $\hbar \omega_B=\hbar \omega_0$. The shift is due 
to the strain modified cyclotron energy.\newline

Indeed, under uniaxial strain, the resonance condition of the $ 0\rightleftarrows \pm 1$ transition in graphene is of the form
(Eqs.~\ref{energyB} and \ref{energyBsim})
\begin{eqnarray}
\hbar \omega_0=\hbar \omega_B^{n=0\pm}&=&\hbar\omega_B\frac {2}3 \frac{\sqrt{w_xw_y}}{at} \left(1-\tilde{w}^2_0\right)^{\frac 34} \nonumber\\
&\sim &\hbar\omega_B\left(1-\frac 13\epsilon\right) \left(1-\frac 34 (0.6\epsilon)^2\right)
\label{reson}
\end{eqnarray}
The magnetic field at which the anticrossing occurs is then $\hbar\omega_B\sim \hbar \omega_0 \left(1+\frac 13\epsilon\right) \left(1+\frac 34 (0.6\epsilon)^2\right)$ which is enhanced (reduced) for
tensile (compressive) deformation.\

In undeformed graphene, the resonance corresponds to $B_0=30.1$ T \cite{Ando}. A strain of $5\%$ is expected to shift this value to
$\tilde{B}_0\sim B_0\left(1+\frac 13\epsilon\right)\sim 30.6$ T. It turns out that a shift of the resonant magnetic field, at which the MPR occurs, could be a signature of the presence of strain in the sample.\

Numerical calculations show that the splitting of the phonon line induced by strain is not significant for small strain values.\

\begin{figure}[hpbt]
\begin{center}
\vspace{0.5cm}

\includegraphics[width=0.9\columnwidth]{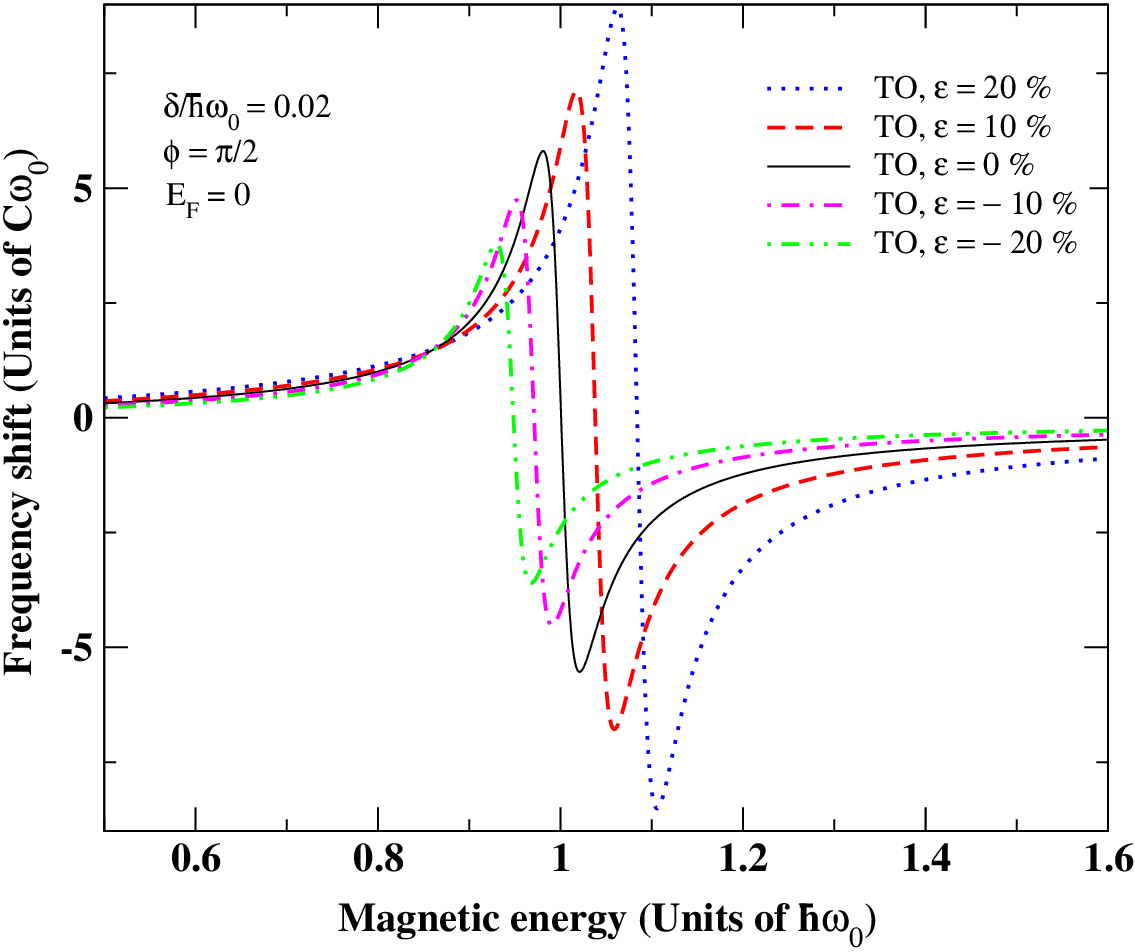}
\end{center}
\caption{Frequency shifts of the TO mode as a function of the magnetic energy $\hbar\omega_B$ 
for different strain values in the undoped graphene and for a disorder amount $\delta=0.02\,\hbar\omega_0$. 
The TO mode is transverse to the strain axis. The thin solid line is the result for the undeformed case.}
\label{shiftstrain}
\end{figure}

For clarity, we consider in the following strain values $|\epsilon|\ge 10\%$ as shown 
in figure \ref{shiftstrain} where we plot the frequency shifts of the TO mode in the undoped regime for different strain values. 
The higher the strain, the larger the shift of the anticrossing position.\

\begin{figure}[hpbt]
\begin{center}
\vspace{0.5cm}

\includegraphics[width=0.9\columnwidth]{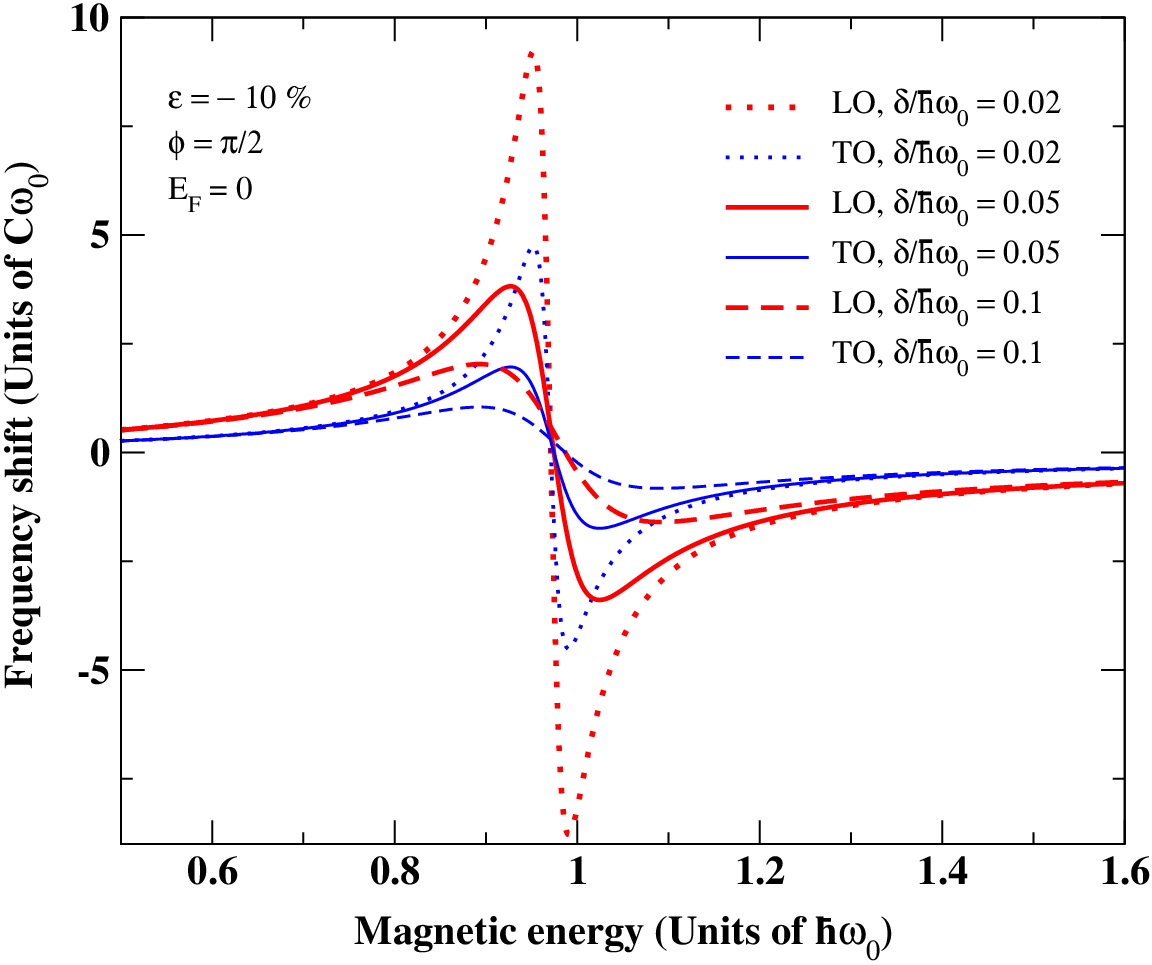}\\

\end{center}
\caption{Frequency shifts of the LO and TO modes as a function of the magnetic energy $\hbar\omega_B$
in the undoped graphene at $\epsilon=-10\%$ and for different disorder amounts.
The shifts and the broadenings of the LO modes, which is along the strain direction, are more pronounced than the corresponding TO modes due to lattice softening along the deformation axis.}
\label{disorder}
\end{figure}

The anti-crossing structure characterizing the phonon frequency line (Figs.~\ref{shiftlow} and \ref{shiftstrain}) is due to the resonance condition 
giving rise to the singular behavior of the phonon self-energy (Eq.~\ref{self}). 
According to this equation, the resonance condition depends on 
the disorder parameter $\delta$. One should then expect a disorder dependence of the MPR lines.\newline
Figure \ref{disorder} shows the frequency shifts of the LO and TO modes for different amounts of disorder
under a compressive deformation $\epsilon=-10\%$. 
The anticrossing is found to be smeared out as the disorder increases as found in the undeformed case \cite{ando2006}.\\

Fig.~\ref{disorder} shows also, that at a given magnetic field, the LO and TO frequency shifts are different as found in zero magnetic field
\cite{Assili}. This feature is ascribed to the dependence of the electron-hole pair production on the strain, leading to
long-lived (damped) phonons along the tensile (compressive) deformation axis\cite{Assili}.\\

\begin{figure}[hpbt]
\begin{center}
\vspace{0.5cm}

\includegraphics[width=1\columnwidth]{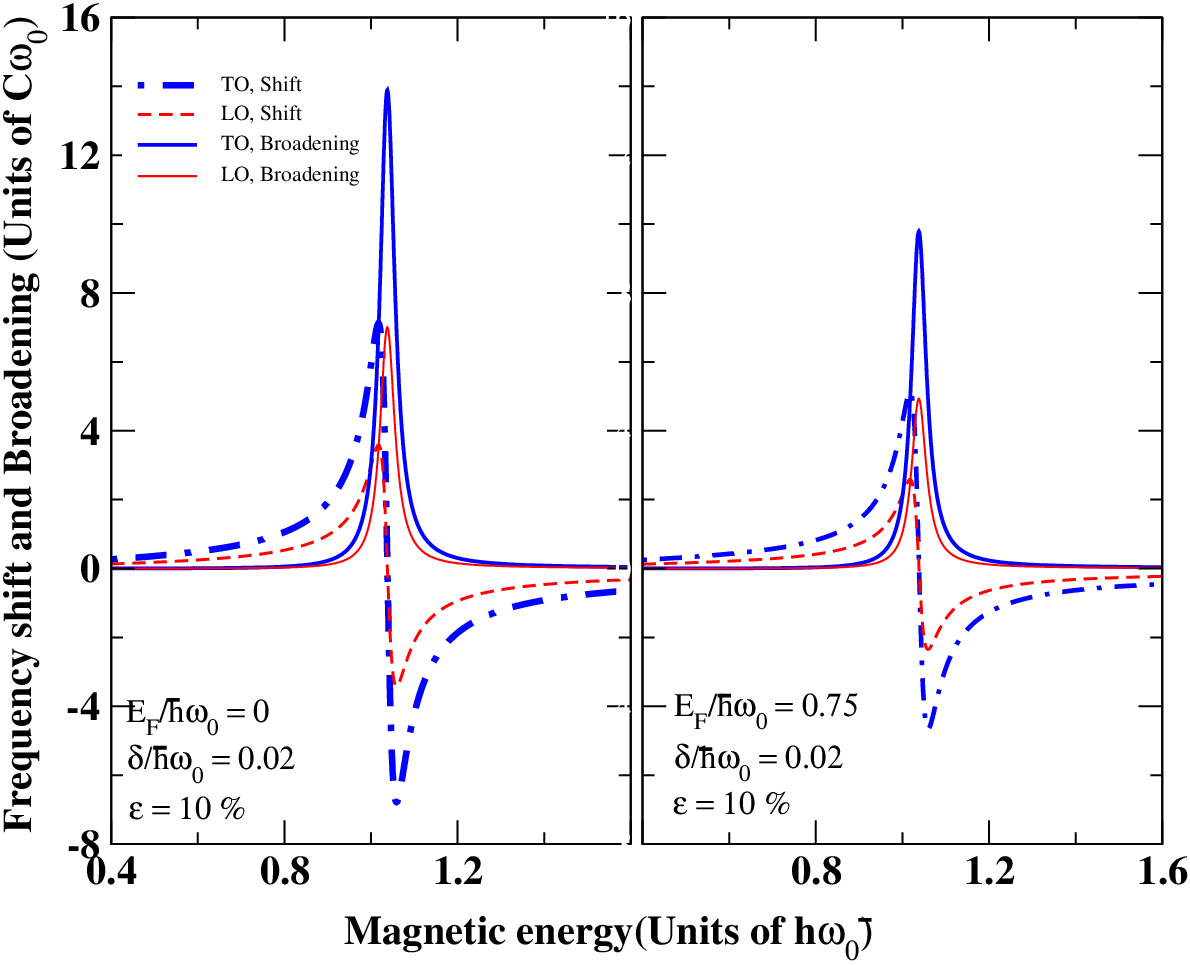}\\

(a)
\vspace{1cm}

\includegraphics[width=1\columnwidth]{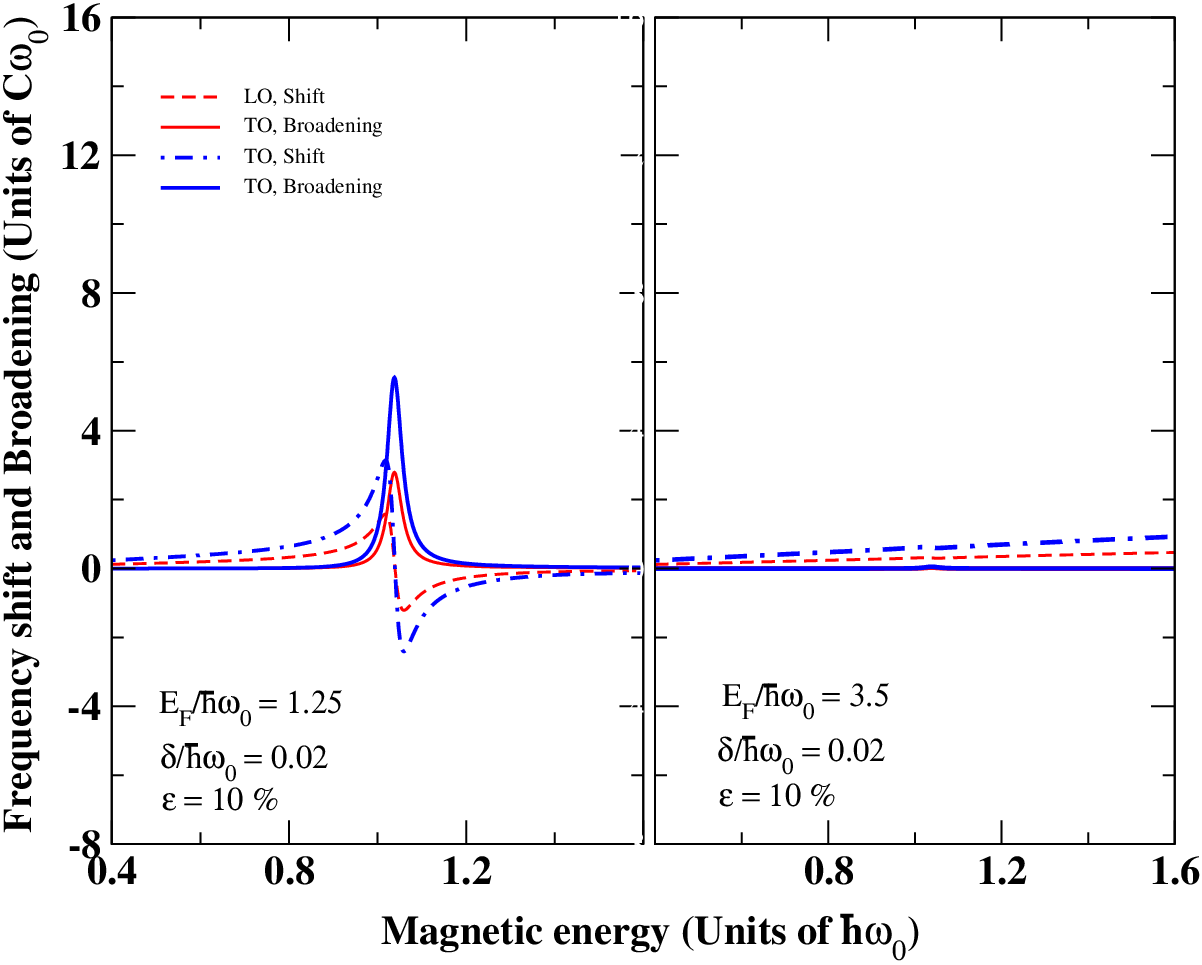}\\

(b)
\end{center}
\caption{Frequency shifts and broadenings of the TO and LO modes as a function of the magnetic 
energy $\hbar\omega_B$ at $\epsilon=10\%$ for different doping levels and for a disorder 
amount of $\delta=0.02\hbar \omega_0$. The LO mode is along the strain axis.}
\label{shiftEF}
\end{figure}

%
Several studies have given evidence of the electron concentration dependence of the MPR \cite{MarkPRL,Yan07,Kossacki,Nano14}. 
This dependence is shown figure \ref{shiftEF} where we plot the frequency shifts and the broadening of the LO and TO modes for different 
Fermi energy value. We denote $\epsilon^{(0)}_F=\sqrt{\frac 32}\hbar \omega_0\sim 1.22\, \hbar \omega_0 $ 
the Fermi energy at which the $n=0$ LL is totally filled in unstrained graphene and $\epsilon^{(1)}_F=\sqrt{2}\left(1+\sqrt{2}\right)\hbar\omega_0\sim 3.4\, \hbar\omega_0$
corresponds to the totally occupied $n=1$ LL \cite{Ando}.\

As shown by Fig.~\ref{shiftEF}, one can distinguish three regimes for the MPR behavior, depending on the Fermi energy: (i) for
$0<E_F < \epsilon^{(0)}_F=\sqrt{\frac 32} \hbar\omega_0 $, where both  $0 \rightleftarrows \pm 1$ transition are active, the anticrossing is well defined with a two strain-induced split lines. These lines are reminiscent of the Raman spectrum obtained by Kashuba and Fal'ko \cite{Kashuba13} at low carrier concentration and for linearly polarized light.\

(ii) For $\epsilon^{(0)}_F<E_F < \epsilon^{(1)}_F\sim 3.4 \,\hbar \omega_0$, the $-1 \rightarrow 0$ transition is blocked by Pauli principle and only the $0 \rightarrow 1$ transition is active.\

The phonon broadening, in this doping regime, is strongly reduced compared to the case of low doping when the $n=0$ is not totally filled.
This reflects the reduction of the number of electron-hole pairs interacting with phonons.\

(iii) For $E_F \geq \epsilon^{(1)}_F$, the $n=1$ LL is totally occupied and the inter-LL transition $0 \rightarrow 1$ is now stopped. The corresponding
MPR disappears. We obtain a single line at the bare phonon frequency $\omega_0$ corresponding to the undeformed lattice. The strain-induced splitting of the phonon line disappears since the latter is due to the coupling of phonons to electron-hole pairs. Such coupling can no more take place since the pair production is blocked by Pauli principle.
Our results are consistent with those obtained by Kashuba and Fal'ko \cite{Kashuba13} in the high doping limit where two field independent Raman lines were
reported for circular polarized light and a single line for linearly polarized light.
The latter shows a relative shift compared to the bare phonon frequency $\omega_0$. This shift, which does not appear in our results, is due to the shear
strain component considered in Ref.~\onlinecite{Kashuba13}. This component gives rise to a degeneracy lifting of the LO and TO phonon modes even in the
absence of the coupling of phonon to electron-hole pairs \cite{Assili}.\\

It is worth mentioning that the interaction of optical phonon with non-relativistic electrons in 2D systems under 
a transverse magnetic field has attracted considerable interest \cite{semicd}.  
This interaction induces an avoided crossing of the LL when the cyclotron resonance frequency matches an appropriate optical phonon 
frequency \cite{Peeters}. 
This effect is substantially different from that studied in the present work where we 
consider the role of electron-phonon coupling on the phonon properties and not on the electronic spectrum.

\subsection{MPR in uniaxial strained graphene under an inplane electric field}
To avoid the collapse of LL, the effective tilt parameter $\tilde{w}_{\xi}(E)$  should satisfy $\tilde{w}_{\xi}(E)<1$ \cite{Lukose,Mark09}. Given the expression of $\tilde{w}_{\xi}(E)$ (Eq.~\ref{tiltE}), the latter condition can be expressed as
\begin{eqnarray}
\tilde{w}_{\xi}(E)\sim |0.6\epsilon-\xi\frac{E}{v_FB}|<1
\label{tiltEapp}
\end{eqnarray}
where we considered the approximate value of $\frac{w_{0x}}{w_x}\sim 0.6 \epsilon$ and put $w_x\sim \hbar v_F$.\

Regarding the small values of the strain amplitude, the condition on the effective tilt parameter $\tilde{w}_{\xi}(E)<1$ reduces to $\frac{E}{v_FB}<1$ as found by Lukose {\it et al.} \cite{Lukose}.
Since the resonance corresponding to $0\leftrightarrows \pm 1$ transitions take place at $B\sim 30$~T and given the graphene average Fermi velocity $v_F\sim 10^6$m.s$^{-1}$, the electric field amplitude should not exceed $10^7$ V$\cdot$m$^{-1}$.\\

\begin{figure}[hpbt]
\begin{center}
\vspace{0.5cm}

\includegraphics[width=0.9\columnwidth]{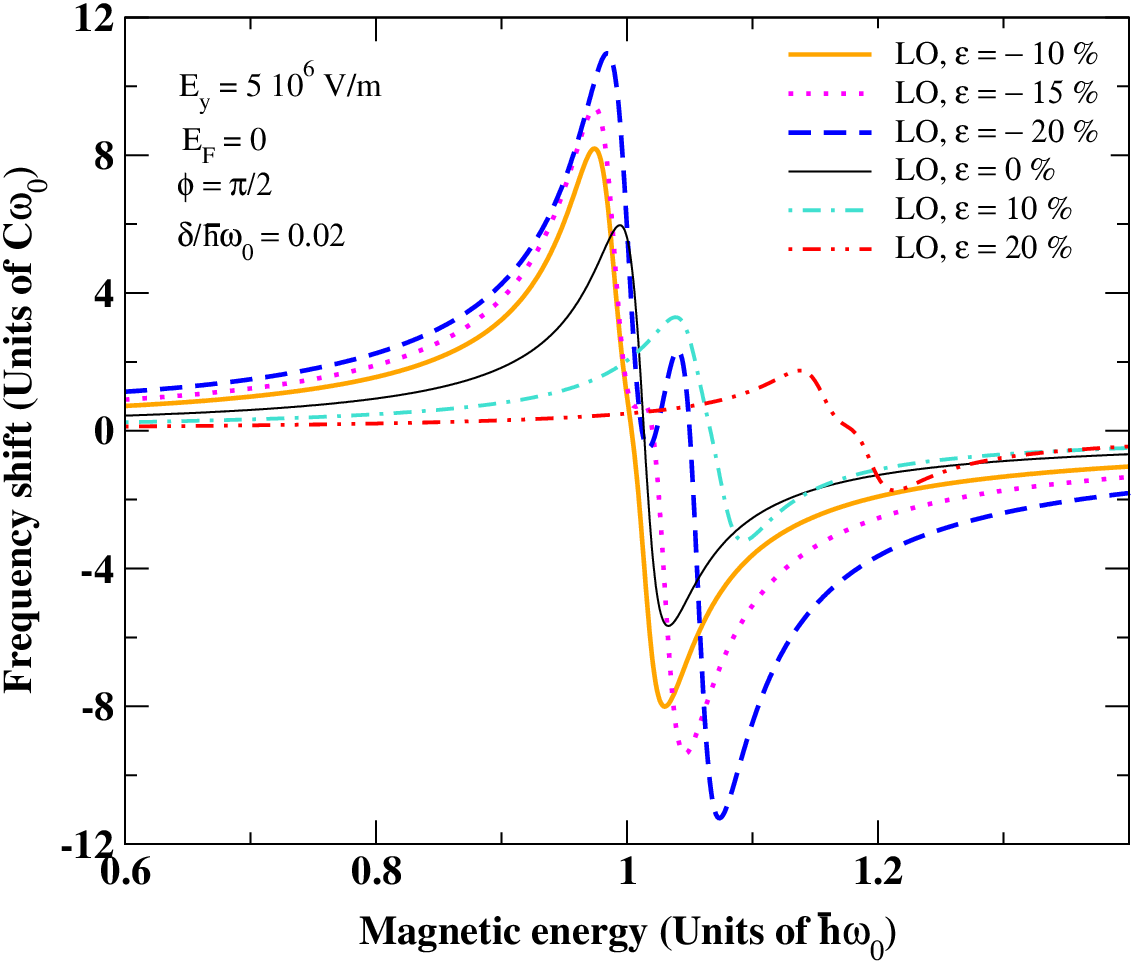}\\

(a)
\vspace{1cm}

\includegraphics[width=0.9\columnwidth]{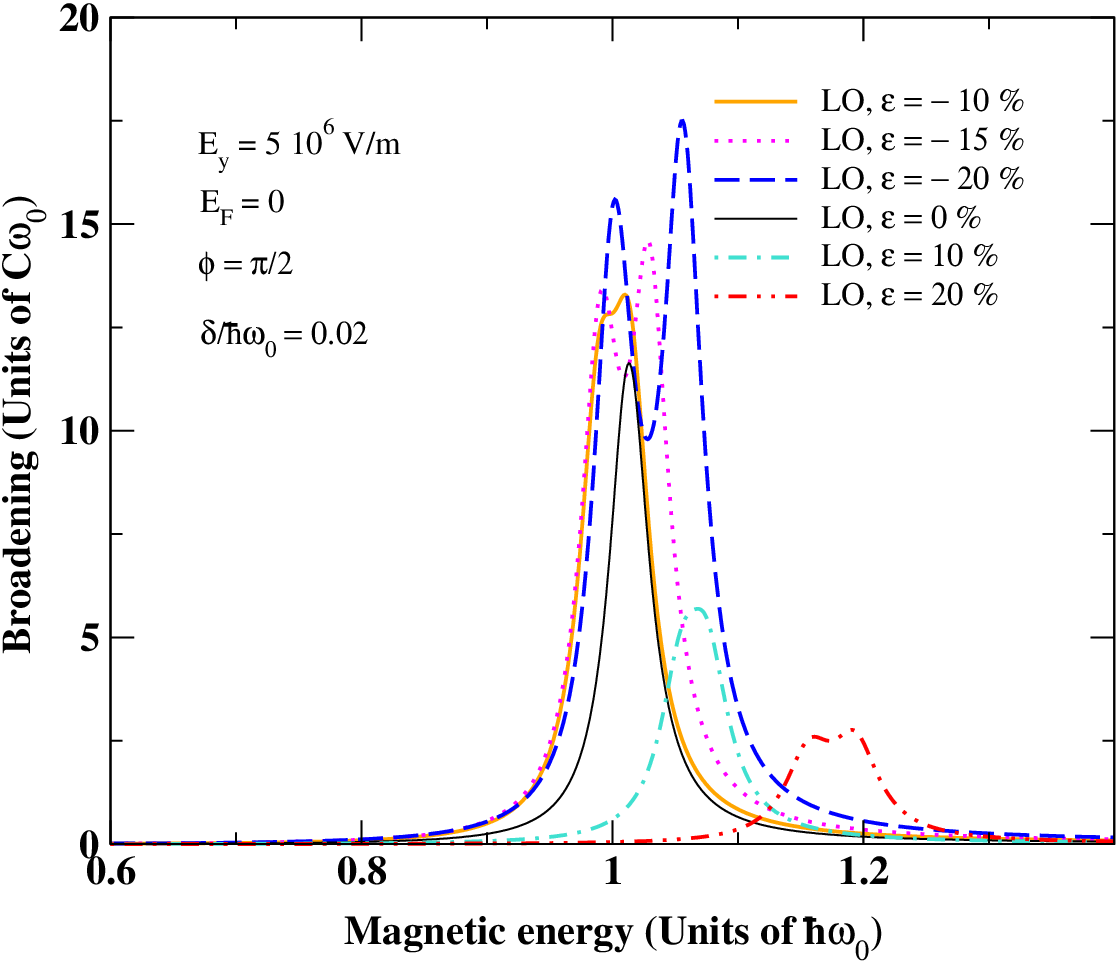}\\

(b)
\end{center}
\caption{Frequency shifts (a) and broadenings (b) of the LO mode as a function of the magnetic energy $\hbar\omega_B$ for different strain strength in the undoped case for a disorder amount $\frac 1{\tau\omega_0}=0.02$ and under a uniform electric field of $5\times 10^6$ V.m$^{-1}$.
The LO mode is along the strain axis. The double peak structure is due to the lifting of the twofold valley degeneracy.}
\label{shift-Ey-strain}
\end{figure}

Figure \ref{shift-Ey-strain} shows the frequency shifts and the broadenings of the LO mode as a function of the magnetic field energy for different strain values in the undoped graphene and under an electric field of $5 \times 10^6$ V.m$^{-1}$.
The shift line exhibits a double peak structure which gets more pronounced as the strain amplitude increases.
According to our numerical results, no double peak is expected at a tensile (compressive) deformation below $\epsilon=15\%$ ($|\epsilon|=10\%$) for a disorder amount of $\delta=0.02\,\hbar \omega_0$.\\

Figure \ref{shift-Ey-strain} shows that, for the chosen disorder amount and doping, the MPR lines exhibit 
a remarkable behavior above the critical strain value.
A double peak structure emerges as a fingerprint of the twofold valley degeneracy lifting induced 
by the effective tilt parameter $\tilde{w}_{\xi}(E)$.
As discussed above, this valley degeneracy lifting is due to a different inter-LL spacing in 
both valleys induced by the inplane electric field (Eq.~\ref{LL-E}). 
Therefore, The inter-LL transitions $0 \leftrightarrows \pm 1$ take place at different resonant magnetic 
fields giving rise to the double resonant MPR lines.\\

It is worth noting that, in the undeformed case (solid line in Fig. \ref{shift-Ey-strain} ), there is no electric field induced double peak. 
The effective tilt parameter reduces, in the absence of strain, to  $\tilde{w}(E)=\frac{\hbar E}{v_F B}$ (Eq.~\ref{tiltEapp})
which is independent of the valley index.
Therefore, the tilt of Dirac cones induced by the uniaxial strain in graphene 
is a substantial ingredient to lift the twofold valley degeneracy in the presence of the inplane electric field. 
The double peak structure of the MPR line emerges as a consequence of the simultaneous action of the Dirac cone tilt and the electric field.\\

Figure \ref{shift-Ey-strain} shows also that, the double peak effect is more pronounced for a compressive deformation. 
This can be understood from the resonance condition, given by Eq.~\ref{reson}, which can be written,
in the presence of the inplane electric field, as:
\begin{eqnarray}
\hbar \omega_0=\hbar\omega^{\xi}_B \left(1-\frac 13\epsilon\right) \left(1-\tilde{w}^2_{\xi}(E)\right)^{\frac 34}
\end{eqnarray}
where we considered the approximate expressions of $w_x$ and $w_y$ (Eq.~\ref{wxapp}). The effective tilt parameter Eq.~\ref{tiltE} can be written as

\begin{eqnarray}
\tilde{w}_{\xi}(E)\sim |0.6\epsilon-\xi\frac{E}{v_FB} \left(1-\frac 23 \epsilon \right)|
\label{tiltapp}
\end{eqnarray}
According to Eq.~\ref{tiltapp}, the larger effective tilt parameter is obtained for a compressive deformation ($\epsilon <0$) 
and in the $\xi=-$ valley. The corresponding resonance is then expected to appear before that ascribed to the $\xi=+$ valley. 
The difference between the valley resonance fields for compressive strain is larger than that obtained for a tensile deformation. 
The double peak structure is then expected to be more pronounced for compressive strain as shown in Fig.~\ref{shift-Ey-strain}.  \\

\begin{figure}[hpbt]
\begin{center}
\vspace{0.5cm}

\includegraphics[width=0.9\columnwidth]{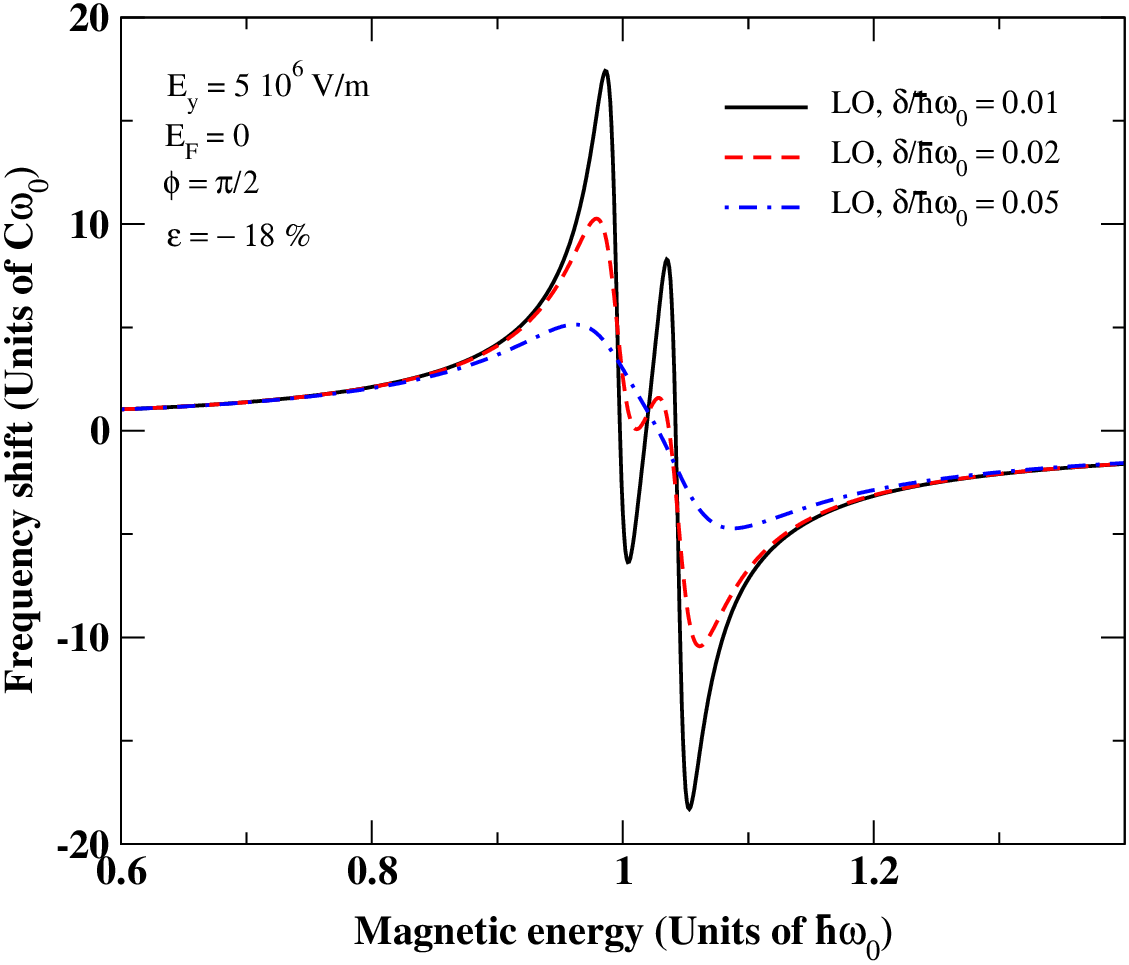}\\

(a)
\vspace{1cm}

\includegraphics[width=0.9\columnwidth]{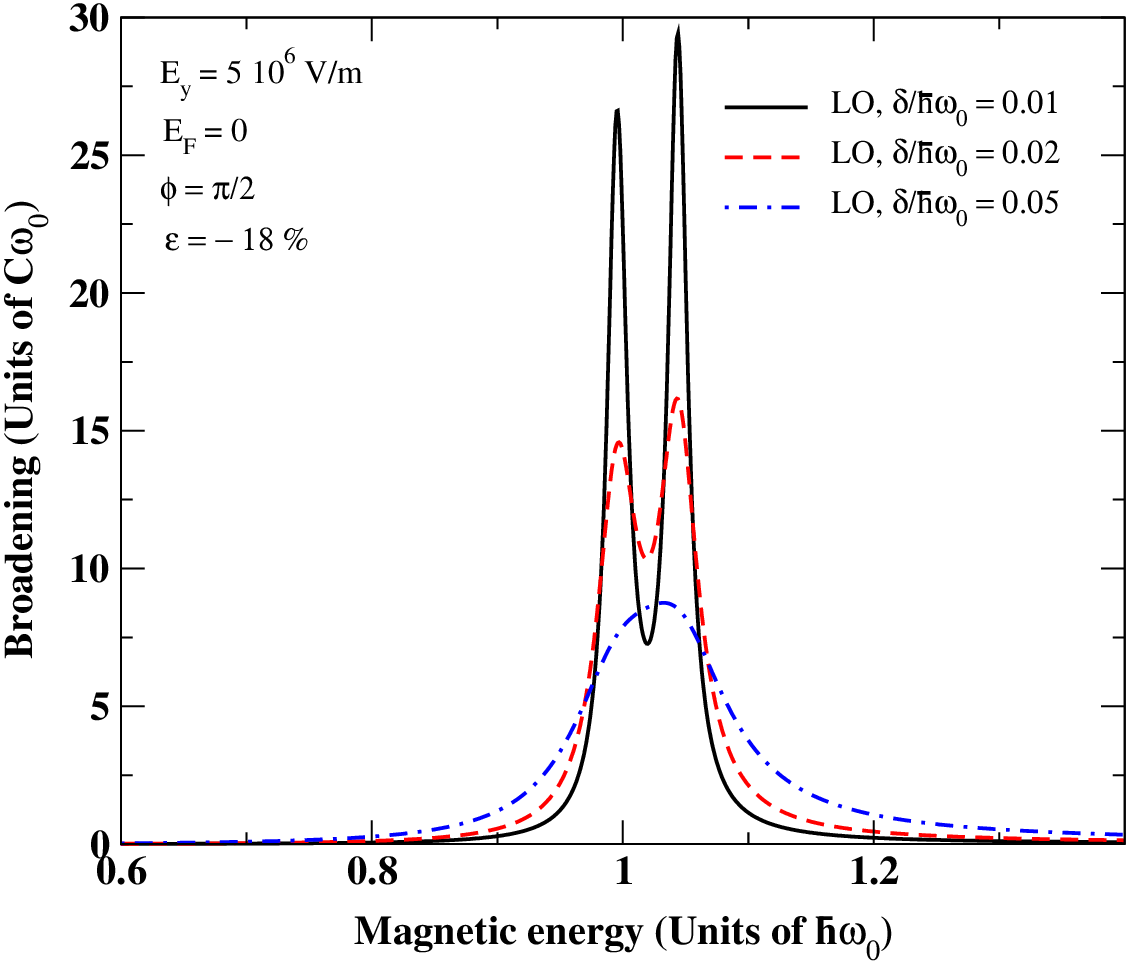}\\

(b)
\end{center}
\caption{Frequency shifts (a) and the broadenings (b) of the LO phonon mode as a function of the magnetic energy $\hbar\omega_B$ for 
compressive deformation of $-18\%$ under a uniform electric field of $5\times 10^6$ V$\cdot$m$^{-1}$ and for different disorder amounts.}
\label{shift-disorder-Ey}
\end{figure}

As discussed in the previous section, the fine structure of the MPR depends on the disorder amount. 
The latter is then expected to be a key parameter for the occurrence of the double peak in the MPR line. 
Figure \ref{shift-disorder-Ey} shows the frequency shifts and the broadenings of the LO phonon mode around the anticrossing for different disorder amount and for an electric field of $5\times 10^6$ V$\cdot$m$^{-1}$.\\

As shown by Fig.~\ref{shift-disorder-Ey} (b), the double peak structure is smeared out by increasing the disorder. 
This effect could be understood from the expression of the phonon self-energy Eq.~\ref{self}. 
The resonance condition reflecting the divergence of the self-energy depends on the disorder parameter $\delta$. 
The latter may, if increased, erase the anticrossing and the relative double peak structure.\\
The double peak separation $\Delta \omega(B)$ is given by $\frac{\Delta \omega(B)}{\omega_0}=\frac{|\omega^{\xi=+}_B-\,\omega^{\xi=-}_B|}{\omega_0}$,
where $\omega^{\xi}_B$ (Eq.~\ref{wBxi}) yields to the singular behavior of the self-energy given by Eq.~\ref{self}. \newline
From this equation, one can deduce a simple condition for the occurrence of the double peak structure as a function of the disorder. 
Such condition could be written as:
\[
 \frac{\Delta \omega(B)}{\omega_0 }> \frac{\delta}{\omega_0 }
\]
The double peak is not smeared out by disorder, as far as this condition is satisfied.\\

\begin{figure}[hpbt]
\begin{center}
\vspace{0.5cm}

\includegraphics[width=0.8\columnwidth]{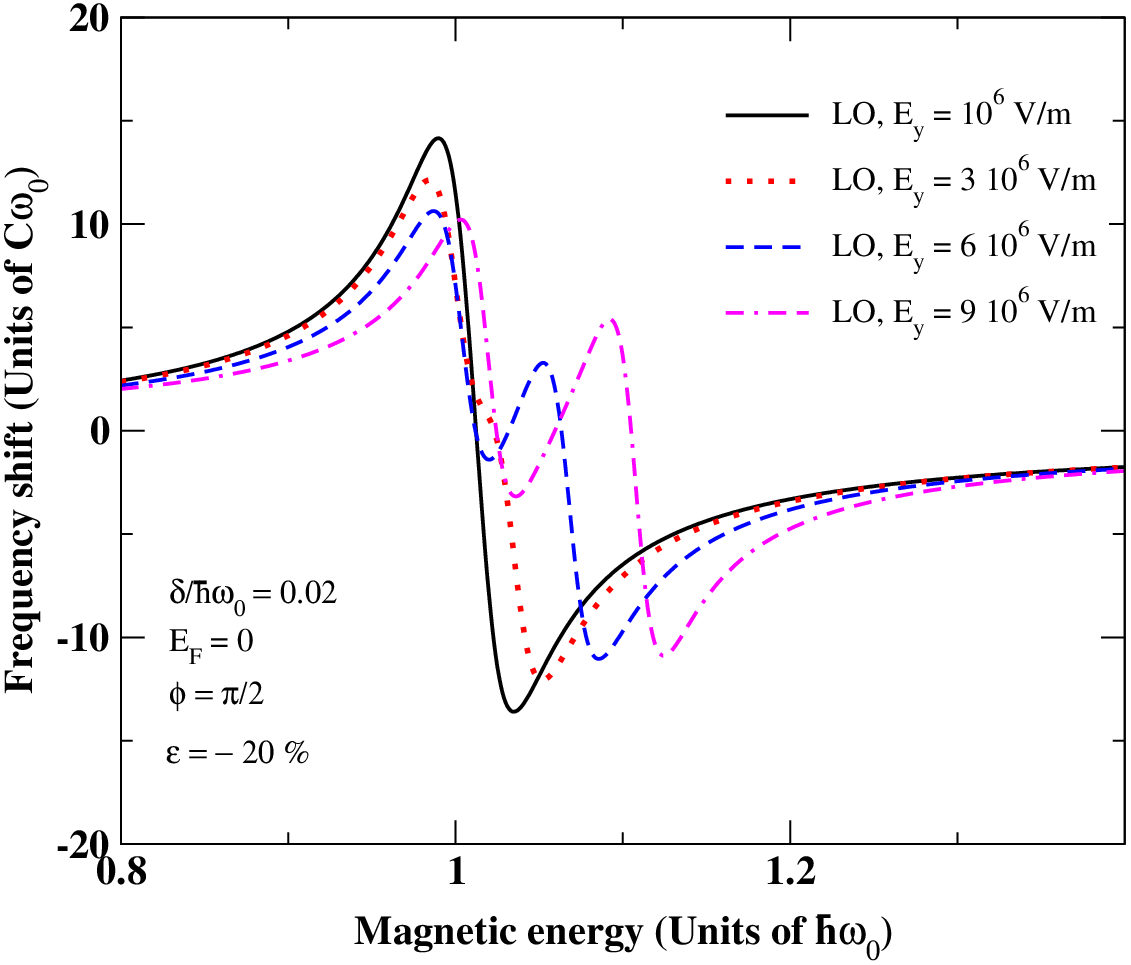}\\

(a)
\vspace{0.8cm}

\includegraphics[width=0.8\columnwidth]{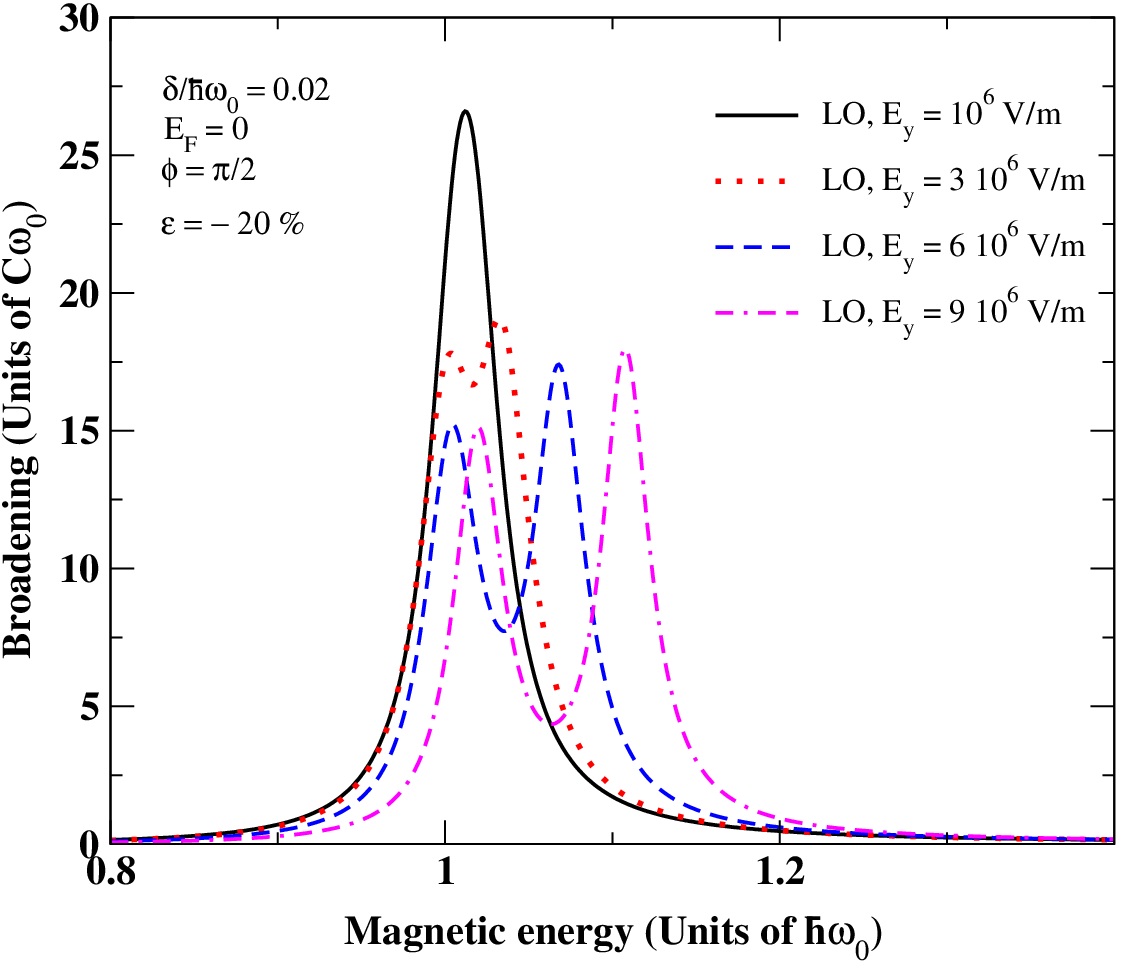}\\

(b)
\end{center}
\caption{Frequency shifts (a) and the broadenings (b) of the LO phonon mode as a function of the magnetic energy $\hbar\omega_B$ for a compressive deformation of $-20\%$ under different electric field  amplitudes and for disorder amount of $\delta/\hbar\omega_0=0.02$.}
\label{Ey}
\end{figure}

The keystone parameter for the occurrence of the double peak structure is the effective tilt of Dirac cones induced by the electric field.
The latter should obey to the condition $\frac E{v_F B}<1$ to avoid the collapse of the LL \cite{Lukose}.
Figure \ref{Ey} shows the MPR double peak structure for different electric field amplitude satisfying the above mentioned condition. 
The double peak structure becomes more pronounced as the electric field increases. This behavior results from the electric field dependence 
of the effective tilt parameter $\tilde{w}_{\xi}(E)$ which is enhanced by increasing the electric field. The higher the electric field, 
the larger the difference between LL spacings in the valleys.\\

\section{Concluding remarks}
We studied the $\Gamma$ point longitudinal (LO) and transverse (TO) optical phonon modes, 
coupled to the inter-LL transition $0\leftrightarrows \pm 1$, around the magneto-phonon resonance field $B\sim $ 30 T in uniaxial strained graphene.
We derived the frequency shifts and the broadenings of the magneto-phonon modes taking into account the coupling of phonons to the strain modified
electronic spectrum. The latter is characterized by tilted Dirac cones with anisotropic velocities.
We found that, disregarding the shear strain component and the strain effect on the phonon dispersion, 
the phonon-electron coupling contributes to the split of the MPR lines and to the shift of the resonant magnetic field.
We also show that the critical doping, and the corresponding filling factor, responsible of the quench of an inter-LL 
transition are strain dependent. This effect may be used to measure the strain amplitude in the sample.
Moreover, we found that, under an inplane electric field, the simultaneous effect of the Dirac cone tilt and 
the electric field, gives rise to a double peak structure in the MPR spectrum of an optical phonon mode $\mu$ ($\mu=$ LO, TO).
This new structure is the signature of the valley degeneracy lifting resulting from different couplings of phonon to electron-hole pairs
in the two valleys, where Landau level spacing are different.
The MPR spectrum of graphene under a uniaxial strain and in the presence of an inplane uniform electric field, 
is then expected to show four peaks, two for each phonon mode.
We show that the double peak structure is substantially dependent on the disorder amount in the lattice, 
on the doping and on the electric field amplitude. We propose that this structure of the MPR spectrum 
corresponding to the $0\leftrightarrows \pm 1$ transitions, could be observed in pristine graphene for a 
uniaxial compressive deformation of $\sim 15\%$ and under an inplane electric field of $5\times 10^6$ V$\cdot$m$^{-1}$ around 
the resonant field $B\sim 30$~T.
\section{Acknowledgment}
We thank J.-N. Fuchs and M.-O. Goerbig for stimulating discussions.
We are indebted to J.-N. Fuchs, D. Basko and M.-O. Goerbig for a critical reading of the manuscript.
This work was partially supported by the National Research Foundation of Korea (NRF) grant funded by the Korea government (MEST) (No. 2012-0008974).
S. H. and M. A. acknowledge the kind hospitality of ICTP (Trieste, Italy) were part of the work  was carried out. S. H. was supported by Simons-ICTP associate fellowship.

\end{document}